\documentclass[11pt,fleqn]{article}
\usepackage[letterpaper,margin=2cm]{geometry}

\usepackage{amsmath,amssymb}
\usepackage{graphicx}
\usepackage{subfig}
\usepackage[numbers]{natbib}
\usepackage{hyperref}
\usepackage{url}
\newcommand{\keywords}[1]{\par\noindent\textbf{Keywords: }#1\par}
\usepackage[ruled]{algorithm2e}
\usepackage{comment}

\hypersetup{colorlinks=true,citecolor=blue,urlcolor=blue,linkcolor=blue}

\title{Analytical Exploration of Spatial Audio Cues: A Differentiable Multi-Sphere Scattering Model}

\author{
Siminfar Samakoush Galougah, Pranav Pulijala, Ramani Duraiswami\thanks{Computational resources were provided by the University of Maryland’s Nexus High-Performance Computing Cluster. We thank Prof. Magnus Wahlberg of SDU, Denmark for discussions and background information on underwater spatial hearing. This work was supported by the Office of Naval Research under Award No.\ N00014-23-1-2086. Data and code are available uppon request.}
 \\
Perceptual Intelligence \& Reality Lab, University of Maryland, College Park, MD, USA
}

\date{ \texttt{\{simin95, pulijala, ramanid\}@umd.edu}} 

\begin{document}
\maketitle


\abstract{A primary challenge in developing synthetic spatial hearing systems, particularly underwater, is accurately modeling sound scattering. Biological organisms achieve 3D spatial hearing by exploiting sound scattering off their bodies to generate location-dependent interaural level and time differences (ITD/ILD). While Head-Related Transfer Function (HRTF) models based on rigid scattering suffice for terrestrial humans, they fail in underwater environments due to the near-impedance match between water and soft tissue. Motivated by the acoustic anatomy of underwater animals, we introduce a novel, analytically derived, closed-form forward model for scattering from a semi-transparent sphere containing two rigid spherical scatterers. This model accurately maps source direction, frequency, and material properties to the pressure field, capturing the complex physics of layered, penetrable structures. Critically, our model is implemented in a fully differentiable setting, enabling its integration with a machine learning algorithm to optimize a cost function for active localization. We demonstrate enhanced convergence for localization under noise using a physics-informed frequency weighting scheme, and present accurate moving-source tracking via an Extended Kalman Filter (EKF) with analytically computed Jacobians. Our work suggests that differentiable models of scattering from layered rigid and transparent geometries offer a promising new foundation for microphone arrays that leverage scattering-based spatial cues over conventional beamforming, applicable to both terrestrial and underwater applications. Our model will be made open source.}


\maketitle

\section{Introduction}
\noindent Biological organisms perform auditory scene analysis in complex environments with just two sensors. They resolve the direction of broadband sources distributed in three-dimensional space, in real time, in the presence of noise, distractors, and reverberation. A fundamental mechanism underlying this ability is the scattering of the incident sound field off the organism's own body, which modifies the signals arriving at the two ears in a location- and frequency-dependent manner. The resulting interaural level and time differences (ILD/ITD) constitute the primary binaural cues for spatial hearing. These cues are combined with connected neural computing to solve various problems in acoustical scene analysis such as localization and tracking.

In terrestrial mammals, the body is largely opaque to sound, and spatial cues are shaped by diffraction around the head and by the elaborate geometry of the outer ear, or pinna. The role of the Head-Related Transfer Function (HRTF), which captures the combined effect of these scattering processes, has been understood for several decades~\cite{Blauert1997}. HRTFs have been measured, computed, and approximated for human listeners, and a number of databases are available. Simplified geometric models have played an important role in understanding the physical mechanisms behind HRTF-based sensing. The rigid-sphere model~\cite{Duda1998} captures diffraction around a single sphere representing the human head, and the ``snowman'' head-and-torso model of~\cite{Gumerov2002-aj} extends this to two rigid spheres using multipole expansion and translation, adding torso reflections and elevation cues~\cite{Gumerov2002-aj, Algazi2002-bh, Gumerov2001-fx}. These models demonstrate that even simple geometries can produce rich spatial information. However, their use in bio-inspired real-time localization or tracking in artificial systems is still in its infancy.

The situation underwater is fundamentally different. Aquatic mammals such as dolphins and toothed whales achieve precise spatial hearing despite lacking external pinnae~\cite{Au1993sonar, Ketten2000}. The acoustic impedance of soft tissue is closely matched to that of seawater, making the body largely transparent to sound. Spatial cues must therefore arise not from diffraction around an opaque obstacle, but from transmission through and scattering within a layered structure of varying acoustic properties. Anatomical and computational studies have shown that the heads of these animals consist of semi-transparent fatty tissues enclosing rigid bony structures, with air sacs acting as acoustic reflectors~\cite{Norris1968, Cranford1996, Ketten2000, Koopman2006, Brill1988, Cranford2010, Cranford2008, Aroyan2001-mb}. Behavioral measurements have confirmed that dolphins exploit interaural time and intensity differences for spatial perception~\cite{Moore1995}. This body of work suggests a general physical principle for underwater spatial sensing: a semi-transparent medium containing internal rigid scatterers can passively shape an incoming sound field into direction-dependent cues.

Abstracting this principle away from any specific biological anatomy, and to develop a simple framework applicable to both underwater and terrestrial settings, we study the scattering from a canonical geometry consisting of a semi-transparent sphere containing two rigid spherical scatterers. We solve the problem analytically using multipole expansions and translations~\cite{GumerovDuraiswami2004, Gumerov2002-np}, obtaining a closed-form forward model that computes the pressure field at any point on the outer surface as a function of frequency, source direction, material properties, and geometric parameters. Implemented in JAX~\cite{bradbury2018jax}, the model is fully differentiable, enabling gradient-based optimization of source direction directly from binaural observations. This extends the tradition of simplified geometric HRTF models in two respects: the outer sphere is permeable rather than rigid, reflecting the semi-transparent nature of biological tissue in water, and the model operates in an acoustic regime where the impedance contrast between the structure and the surrounding medium is small.

We demonstrate the model through several applications. We first analyze the HRTF, ILD, and ITD patterns produced by the geometry, confirming that it generates directionally dependent binaural cues. We then formulate source localization as a computing mechanism connected to sensing, that performs gradient-based optimization of a joint ILD-ITD loss. We evaluate robustness to measurement noise and apply an extended Kalman filter (EKF) for tracking a moving source, using the differentiable model to compute the required Jacobians.





The remainder of this paper is organized as follows. Section~\ref{sec:2} reviews the relevant background. Section~\ref{sec:TwoSLinSmt} presents the problem formulation and analytical solution. Section~\ref{sec:spatialcues} defines the spatial cues and the source localization framework. Section~5 presents simulation results, and Section~6 concludes the paper.

 
\section{\label{sec:2} Background}

\subsection{Classical HRTF Modeling}

\noindent The HRTF characterizes how an incoming sound field is modified by scattering off a listener's head, torso, and pinnae before reaching the eardrums~\cite{Blauert1997}. For a source at position $\mathbf{r}_s$ and a receiver at position $\mathbf{r}$ on or near the head, the HRTF is defined as the ratio of the sound pressure in the presence of the scatterer to the pressure that would exist at the same point in free field:
\begin{equation}
H(\mathbf{r}, \mathbf{r}_s, \omega) 
= \frac{p_{\mathrm{total}}(\mathbf{r}, \omega)}{p_{\mathrm{free}}(\mathbf{r}, \omega)},
\label{eq:hrtf_def}
\end{equation}
where $\omega$ is the angular frequency. The HRTF is complex-valued, encoding both amplitude and phase modifications that depend on frequency and the relative geometry of source and receiver. When evaluated at the two ears, the HRTF pair gives rise to ILD and ITD, which are the primary cues for spatial hearing.

Spherical modeling of the HRTF began with the rigid-sphere approximation. Duda and Martens~\cite{Duda1998} showed that a single rigid sphere captures some essential features of human head diffraction, producing frequency-dependent ILD and ITD that match measurements reasonably well for azimuthal localization. The ``snowman'' head-and-torso simulator~\cite{Algazi2002-bh} extends this to two rigid spheres, capturing torso reflections and improving elevation cues. Several simplified geometric HRTF models have been developed along these lines~\cite{Gumerov2002-aj, Gumerov2001-fx}, demonstrating that even elementary geometries can produce spatially rich binaural behavior.

While these rigid-sphere models are effective for terrestrial hearing, where the head is largely opaque to sound, they are not adequate for underwater settings. When the acoustic impedance of the scattering body is close to that of the surrounding medium, most of the incident energy is transmitted rather than reflected, and a rigid boundary condition no longer applies.

\subsection{Multi-Spherical and Penetrable Scattering Models}

\noindent A sphere is termed \textit{semi-transparent} when its acoustic impedance differs from, but is not vastly larger than, that of the surrounding medium. Sound incident on such a sphere is partially reflected and partially transmitted, with the balance governed by the impedance contrast. At the interface, both the acoustic pressure and the normal component of particle velocity must be continuous, leading to transmission conditions rather than the rigid (Neumann) or soft (Dirichlet) boundary conditions used for opaque scatterers. In the limiting case where the interior impedance matches the exterior exactly, the sphere becomes fully transparent and produces no scattering.

To model structures with multiple layers of different acoustic properties, researchers have extended the single-sphere framework to penetrable and multi-layered geometries~\cite{Cai2011MultilayerSphere}. For concentric spherical shells, continuity of pressure and normal velocity at each interface couples the modal coefficients across layers, yielding a system solvable via transfer matrices~\cite{pierce2019acoustics}.

When the inner spheres are not concentric with the outer shell, the scattered field from each sphere must be re-expressed in the coordinate system of the others. This is accomplished through translation of wave function expansions using re-expansion coefficients~\cite{GumerovDuraiswami2004}. Two types of re-expansion arise: singular-to-regular $(S|R)$ coefficients, which translate the scattered (outgoing) field of one sphere into a regular (incoming) expansion about another center, and regular-to-regular $(R|R)$ coefficients, which translate a regular field between centers. These re-expansion operations are exact and preserve the modal structure of the fields.

Gumerov and Duraiswami~\cite{Gumerov2002-np} developed a framework for computing scattering from multiple rigid spheres using multipole re-expansion, which forms the mathematical foundation for the present work. Their approach solves for the scattered field coefficients of each sphere simultaneously by coupling the boundary conditions through translation operators. In this paper, we extend this framework to the case where rigid spheres are enclosed within a semi-transparent sphere, requiring both the rigid boundary conditions on the inner spheres and transmission conditions on the outer sphere to be satisfied simultaneously.

\subsection{Differentiable Acoustic Simulation}

\noindent Numerical methods such as the boundary element method (BEM), finite element method (FEM), and fast multipole method (FMM) can compute HRTFs for arbitrary geometries with high accuracy~\cite{kahana1999numerical}. However, these methods are computationally expensive, particularly at high frequencies, and do not readily provide gradients of the output with respect to geometric or material parameters. This limits their use in optimization or inverse problems where one wishes to tune the physical design to achieve desired acoustic properties.

Automatic differentiation (AD) frameworks such as JAX~\cite{bradbury2018jax} and PyTorch, primarily developed for deep learning, offer an alternative approach for physics-based optimization. When the forward model is implemented in such a framework, gradients of any scalar loss function with respect to all input parameters are computed exactly and efficiently via the chain rule, without finite differences or adjoint derivations. This paradigm has been exploited in a growing body of work, including differentiable physics simulation~\cite{oktay2024translating} and differentiable microphone arrays~\cite{galougah2024}. In the context of the present work, implementing the multipole scattering model in JAX makes the entire forward map (from geometric parameters, material properties, and source direction to the pressure at the sensor locations) fully differentiable. This enables gradient-based optimization of source direction for localization, as well as sensitivity analysis and, in principle, inverse design of the physical structure.

\section{Methodology}

\subsection{Problem Geometry and Governing Equations}
\label{sec:TwoSLinSmt}

\begin{figure}[htb]
    \centering
    \includegraphics[width=0.8\linewidth]{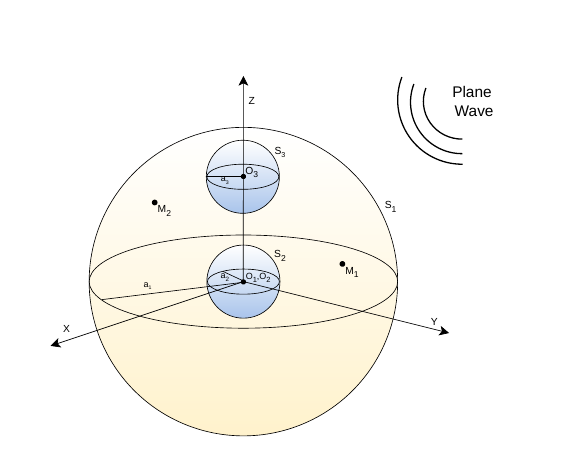}
    \caption{Problem geometry: two rigid spheres $S_2$ and $S_3$ inside a semi-transparent sphere $S_1$. Spheres $S_1$ and $S_2$ are concentric. The exterior medium has density $\rho_o$ and sound speed $c_o$; the interior medium has density $\rho_i$ and sound speed $c_i$.}
    \label{fig:1SL2TS}
\end{figure}

We consider the scattering of a time-harmonic acoustic field from a configuration of three spheres: a semi-transparent (penetrable) outer sphere containing two rigid inner spheres, all immersed in a homogeneous fluid medium. The geometry is depicted in Fig.~\ref{fig:1SL2TS}.

Let $S_1$ denote the semi-transparent outer sphere with radius $a_1$ and center $\mathbf{r}'_1$, and let $S_2$ and $S_3$ denote the two rigid inner spheres with radii $a_2$, $a_3$ and centers $\mathbf{r}'_2$, $\mathbf{r}'_3$, respectively. The exterior medium (outside $S_1$) has density $\rho_o$ and sound speed $c_o$, while the interior medium (inside $S_1$ but outside $S_2$ and $S_3$) has density $\rho_i$ and sound speed $c_i$. The corresponding wavenumbers are $k_o = \omega/c_o$ and $k_i = \omega/c_i$, where $\omega$ is the angular frequency. For simplicity, we take $S_1$ and $S_2$ to be concentric, so that $\mathbf{r}'_2 = \mathbf{r}'_1$, while $S_3$ is offset from the common center.

We introduce two coordinate systems: $(O_1 x_1 y_1 z_1)$ centered at $\mathbf{r}'_1$ (which also serves as the coordinate system for $S_2$), and $(O_3 x_3 y_3 z_3)$ centered at $\mathbf{r}'_3$. Since $S_1$ and $S_2$ share the same center, re-expansion between their coordinate systems is the identity; only translations between $O_1$ and $O_3$ are required.

\subsubsection{Governing equation}

The acoustic pressure in each region satisfies the wave equation. In what follows we deal with the Fourier transform of the pressure, $\psi$, given by
$$
\psi\left(  \mathbf{r,}\omega\right)  =\int_{-\infty}^{\infty}e^{i\omega
t}p\left(  \mathbf{r,}t\right)  \text{d}t,
$$
which satisfies the Helmholtz equation. In the exterior region (outside $S_1$):
\begin{equation}
\nabla^2 \psi_o + k_o^2 \psi_o = 0,
\label{eq:helmholtz_ext}
\end{equation}
and in the interior region (inside $S_1$, outside $S_2$ and $S_3$):
\begin{equation}
\nabla^2 \psi_{\mathrm{internal}} + k_i^2 \psi_{\mathrm{internal}} = 0.
\label{eq:helmholtz_int}
\end{equation}

\subsubsection{Boundary conditions}

On the surfaces of the rigid spheres $S_2$ and $S_3$, the normal component of particle velocity vanishes, giving the Neumann conditions:
\begin{align}
|\mathbf{r}-\mathbf{r}'_2| = a_2:\quad
  \frac{\partial \psi_{\mathrm{internal}}}{\partial n_2} &= 0,
  \label{eq:bc_S2}\\[4pt]
|\mathbf{r}-\mathbf{r}'_3| = a_3:\quad
  \frac{\partial \psi_{\mathrm{internal}}}{\partial n_3} &= 0,
  \label{eq:bc_S3}
\end{align}
where the normal derivatives are taken in the direction of the outward normals $\mathbf{n}_q$ on the surfaces $S_q$ for $q = 2, 3$.

On the surface of the semi-transparent sphere $S_1$, both the normal particle velocity and the pressure must be continuous across the interface. These transmission conditions are:
\begin{align}
|\mathbf{r}-\mathbf{r}'_1| = a_1:\quad
  \frac{\partial \psi_{\mathrm{internal}}}{\partial n_1}
  &= \frac{\partial \psi_o}{\partial n_1},
  \label{eq:bc_S1_vel}\\[4pt]
  \omega_i\rho_i\;\psi_{\mathrm{internal}}
  &= \omega_o\rho_o\;\psi_o,
  \label{eq:bc_S1_pres}
\end{align}
where $\omega_i$ and $\omega_o$ are the angular frequencies inside and outside $S_1$, respectively. These conditions ensure continuity of normal velocity (Eq.~\ref{eq:bc_S1_vel}) and continuity of pressure scaled by the medium impedance (Eq.~\ref{eq:bc_S1_pres}).

In addition, the scattered field must satisfy the Sommerfeld radiation condition at infinity, ensuring that it represents outgoing waves.

\subsubsection{Field decomposition}

The total field in the exterior region is decomposed as the sum of the known incident field and the scattered field due to $S_1$:
\begin{equation}
\psi_o(\mathbf{r})
= \psi_{\mathrm{in}}(\mathbf{r})
  + \psi_{1,\mathrm{scat}}(\mathbf{r}).
\label{eq:outside_decomp}
\end{equation}
The total field in the interior region consists of the transmitted incident field inside $S_1$ and the fields scattered by the two rigid spheres:
\begin{equation}
\psi_{\mathrm{internal}}(\mathbf{r})
= \psi_{1,\mathrm{inside}}(\mathbf{r})
  + \psi_{2,\mathrm{scat}}(\mathbf{r})
  + \psi_{3,\mathrm{scat}}(\mathbf{r}).
\label{eq:inside_decomp}
\end{equation}

\subsubsection{Multipole representation}

We represent each field component using spherical multipole expansions. In spherical coordinates $\mathbf{r} = (r, \theta, \phi)$, the regular and singular basis functions are defined as
\[
R_l^s(\mathbf{r}) = j_l(kr)\,Y_l^s(\theta,\phi),
\qquad
S_l^s(\mathbf{r}) = h_l(kr)\,Y_l^s(\theta,\phi),
\]
where $j_l$ is the spherical Bessel function of the first kind, $h_l$ is the spherical Hankel function of the first kind, and $Y_l^s$ is the spherical harmonic of degree $l$ and order $s$, defined as~\cite{b32}
\begin{equation}
Y_l^{s}(\theta,\phi)
= \sqrt{\frac{2l+1}{4\pi}}
  \sqrt{\frac{(l-|s|)!}{(l+|s|)!}}\;
  P_l^{|s|}(\cos\theta)\;e^{js\phi},
\label{eq:Ylm}
\end{equation}
with $P_l^{|s|}$ the associated Legendre function. Regular basis functions $R_l^s$ are finite at the origin and represent incident or transmitted fields, while singular basis functions $S_l^s$ satisfy the radiation condition and represent scattered (outgoing) fields.

The incident and transmitted fields are expanded in regular basis functions about $O_1$:
\begin{align}
\psi_{\mathrm{in}}(\mathbf{r})
  &= \sum_{l=0}^{\infty}\sum_{s=-l}^{l}
     E_l^{s}\, R_l^s(\mathbf{r}_1),
\label{eq:psi_in}\\[4pt]
\psi_{1,\mathrm{inside}}(\mathbf{r})
  &= \sum_{l=0}^{\infty}\sum_{s=-l}^{l}
     A_l^{(1)s}\, R_l^s(\mathbf{r}_1),
\label{eq:psi_inside}
\end{align}
and the scattered fields are expanded in singular basis functions in their respective coordinate systems:
\begin{align}
\psi_{1,\mathrm{scat}}(\mathbf{r})
  &= \sum_{l=0}^{\infty}\sum_{s=-l}^{l}
     B_l^{(1)s}\, S_l^s(\mathbf{r}_1),
\label{eq:psi_1scat}\\[4pt]
\psi_{2,\mathrm{scat}}(\mathbf{r})
  &= \sum_{l=0}^{\infty}\sum_{s=-l}^{l}
     C_l^{(1)s}\, S_l^s(\mathbf{r}_1),
\label{eq:psi_2scat}\\[4pt]
\psi_{3,\mathrm{scat}}(\mathbf{r})
  &= \sum_{l=0}^{\infty}\sum_{s=-l}^{l}
     D_l^{(3)s}\, S_l^s(\mathbf{r}_3).
\label{eq:psi_3scat}
\end{align}
The superscripts on the coefficients denote the coordinate system in which they are defined. The unknown coefficients to be determined are $A_l^{(1)s}$ (transmitted field inside $S_1$), $B_l^{(1)s}$ (scattered field from $S_1$), $C_l^{(1)s}$ (scattered field from $S_2$), and $D_l^{(3)s}$ (scattered field from $S_3$).

\subsubsection{Incident field}

For a monopole point source of strength $Q$ located at $\mathbf{r}_s$, the incident field is
\begin{equation*}
\psi_{\mathrm{in}}(\mathbf{r})
= Q\,G_k(\mathbf{r}-\mathbf{r}_s)
= Q\,\frac{\exp(ik_o|\mathbf{r}-\mathbf{r}_s|)}{4\pi|\mathbf{r}-\mathbf{r}_s|},
\end{equation*}
where $G_k$ is the free-space Green's function. Expanding about $O_1$ in multipoles gives~\cite{GumerovDuraiswami2004}
\begin{equation*}
\psi_{\mathrm{in}}(\mathbf{r})
= Q\,ik_o \sum_{l=0}^{\infty}\sum_{s=-l}^{l}
  S_l^{-s}(\mathbf{r}_s)\,R_l^s(\mathbf{r}),
\end{equation*}
so that $E_l^{s} = Q\,ik_o\,S_l^{-s}(\mathbf{r}_s)$.

For a plane wave incident from direction $(\theta_s, \phi_s)$, the incident field is $\psi_{\mathrm{in}}(\mathbf{r}) = e^{i\mathbf{k}\cdot\mathbf{r}}$, which expands as
\[
\psi_{\mathrm{in}}(\mathbf{r})
= \sum_{l=0}^{\infty} i^l(2l+1)\,j_l(k_o r)\,P_l(\cos\theta),
\]
yielding the coefficients $E_l^{s} = 4\pi\,i^l\,\overline{Y_l^{s}(\theta_s,\phi_s)}$ under the Condon--Shortley phase convention, where the overline denotes complex conjugation. In the numerical experiments presented in this paper, we use the plane-wave incident field.

\subsection{Analytical Solution}
\label{sec:AnSol}

To solve the coupled boundary-value problem defined in Section~\ref{sec:TwoSLinSmt}, we substitute the multipole expansions into the boundary conditions on each sphere. Because the scattered field of $S_3$ is expressed in the coordinate system $O_3$ while the boundary conditions on $S_1$ and $S_2$ are most naturally enforced in $O_1$, we must first re-expand $\psi_{3,\mathrm{scat}}$ about $O_1$ using the singular-to-regular translation:
\begin{equation}
\psi_{3,\mathrm{scat}}(\mathbf{r})
= \sum_{l=0}^{\infty}\sum_{s=-l}^{l}
  D_l^{(1)s}\,R_l^s(\mathbf{r}_1),
\label{eq:reexpand_D_to_O1}
\end{equation}
where
\begin{equation}
D_l^{(1)s}
= \sum_{m=|s|}^{\infty}
  (S|R)_{ml}^{s}(\mathbf{r}'_{31})\;D_m^{(3)s}.
\label{eq:D_translation}
\end{equation}
Here, $\mathbf{r}'_{tz} = \mathbf{r}'_z - \mathbf{r}'_t$ denotes the translation vector from the center of sphere $S_t$ to the center of sphere $S_z$, and $(S|R)_{ml}^s$ are the singular-to-regular coaxial re-expansion coefficients~\cite{GumerovDuraiswami2004}, which translate an outgoing (singular) expansion about one center into a regular expansion about another.

\subsubsection{Boundary conditions on $S_1$}

Substituting the multipole expansions into the velocity continuity condition~\eqref{eq:bc_S1_vel} and exploiting the orthogonality of the spherical harmonics yields, for each degree $l$ and order $s$:
\begin{multline}
A_l^{(1)s}\,k_i\,j_l'(k_i a_1)
+ C_l^{(1)s}\,k_i\,h_l'(k_i a_1)
+ D_l^{(1)s}\,k_i\,j_l'(k_i a_1)\\
- B_l^{(1)s}\,k_o\,h_l'(k_o a_1)
= E_l^{s}\,k_o\,j_l'(k_o a_1).
\label{eq:bc1_vel}
\end{multline}
Similarly, from the pressure continuity condition~\eqref{eq:bc_S1_pres}, defining $d = \omega_i \rho_i / (\omega_o \rho_o)$:
\begin{multline}
d\!\left[
  A_l^{(1)s}\,j_l(k_i a_1)
  + C_l^{(1)s}\,h_l(k_i a_1)
  + D_l^{(1)s}\,j_l(k_i a_1)
\right]\\
- B_l^{(1)s}\,h_l(k_o a_1)
= E_l^{s}\,j_l(k_o a_1).
\label{eq:bc1_pres}
\end{multline}

\subsubsection{Boundary condition on $S_2$}

Since $S_2$ is concentric with $S_1$, all fields are already expressed in the common coordinate system $O_1$. Applying the Neumann condition~\eqref{eq:bc_S2} gives:
\begin{equation}
A_l^{(1)s}\,k_i\,j_l'(k_i a_2)
+ C_l^{(1)s}\,k_i\,h_l'(k_i a_2)
+ D_l^{(1)s}\,k_i\,j_l'(k_i a_2) = 0.
\label{eq:bc2}
\end{equation}

\subsubsection{Boundary condition on $S_3$}

To enforce the Neumann condition on $S_3$, we need the interior field expressed in the coordinate system $O_3$. The transmitted field $\psi_{1,\mathrm{inside}}$ and the scattered field $\psi_{2,\mathrm{scat}}$, both defined about $O_1$, are re-expanded about $O_3$ using the appropriate translation operators:
\begin{multline}
\!\!\!\!\psi_{\mathrm{internal}}(\mathbf{r}_3)
= \psi_{1,\mathrm{inside}}(\mathbf{r}_3)
  + \psi_{2,\mathrm{scat}}(\mathbf{r}_3)
  + \psi_{3,\mathrm{scat}}(\mathbf{r}_3)\!= \\
 \!\!\sum_{l=0}^{\infty}\sum_{s=-l}^{l}\!\!
  \Big[
A_l^{(3)s}\,R_l^s(k_i\mathbf{r}_3) + 
C_l^{(3)s}\,R_l^s(k_i\mathbf{r}_3) +
D_l^{(3)s}\,S_l^s(k_i\mathbf{r}_3)
  \Big],
\label{eq:internal_O3}
\end{multline}
where the translated coefficients are
\begin{align}
A_l^{(3)s}
  &= \sum_{m=|s|}^{\infty}
     (R|R)_{ml}^{s}(\mathbf{r}'_{13})\;A_m^{(1)s},
\label{eq:A_translation}\\[4pt]
C_l^{(3)s}
  &= \sum_{m=|s|}^{\infty}
     (S|R)_{ml}^{s}(\mathbf{r}'_{13})\;C_m^{(1)s}.
\label{eq:C_translation}
\end{align}
Here, $(R|R)_{ml}^s$ are the regular-to-regular coaxial re-expansion coefficients, which translate a regular expansion from one center to another~\cite{GumerovDuraiswami2004}. Note that $\psi_{2,\mathrm{scat}}$, being a singular expansion about $O_1$, is translated using $(S|R)$ coefficients, converting it to a regular expansion about $O_3$.

Applying the Neumann condition~\eqref{eq:bc_S3} in the $O_3$ coordinate system gives:
\begin{equation}
A_l^{(3)s}\,k_i\,j_l'(k_i a_3)
+ C_l^{(3)s}\,k_i\,j_l'(k_i a_3)
+ D_l^{(3)s}\,k_i\,h_l'(k_i a_3) = 0.
\label{eq:bc3}
\end{equation}

Further details of solving linear systems mentioned in the above subsections and finding coefficients $A, ~B,~ C, ~\text{and}~ D$ are provided in Appendix~\ref{linear_system}.

The entire forward model from geometry and material parameters to the pressure at any evaluation point is implemented in JAX~\cite{bradbury2018jax}, making it fully differentiable with respect to all inputs via automatic differentiation.

\section{Spatial Cues}
\label{sec:spatialcues}

Given the solution to the scattering problem, we can evaluate the acoustic pressure at any point on the surface of $S_1$. We place two sensors at positions $M_L$ and $M_R$ on the outer sphere, representing the left and right receivers. The direction-dependent differences between the signals at these two points constitute the binaural spatial cues used for source localization.
\subsection{Definitions and Formulas}
The HRTF at a point $M$ on the surface of $S_1$ is defined as the ratio of the total exterior pressure at $M$ to the free-field pressure from the source:
\begin{equation}
H(M) = \frac{\psi_o(M)}{Q\,G_k(r_s)},
\qquad M \in S_1,
\label{eq:HRTF_def}
\end{equation}
where $Q$ is the source strength and $G_k(r_s)$ is the free-space Green's function evaluated at the source distance. The HRTF is complex-valued and depends on frequency and the angular position of the source relative to the scatterer.

The ILD quantifies the difference in received amplitude between the two sensors, expressed in decibels:
\begin{equation}
\mathrm{ILD}(f)
= 20\,\log_{10}\!\frac{|H(M_R, f)|}{|H(M_L, f)|}.
\label{eq:ILD}
\end{equation}
A positive ILD indicates that the source is closer to or less shadowed from the right sensor; the magnitude and sign vary with frequency and source direction due to the frequency-dependent scattering produced by the geometry.

The ITD quantifies the difference in arrival time between the two sensors. We compute a frequency-dependent ITD from the interaural phase difference:
\begin{equation}
\mathrm{ITD}(f)
= \frac{\angle H(M_R, f) - \angle H(M_L, f)}{2\pi f},
\label{eq:ITD_phase}
\end{equation}
where $\angle(\cdot)$ denotes the unwrapped phase. This phase-based definition yields a frequency-dependent ITD that captures the dispersive effects of scattering through the layered structure, which is particularly relevant in the present geometry where the interior medium has a different sound speed from the exterior.

Together, the ILD and ITD encode the source direction in complementary ways: the ILD is most informative at higher frequencies where the scatterer dimensions are large relative to the wavelength, while the ITD provides robust directional information at lower frequencies. The forward model described in Sections~\ref{sec:TwoSLinSmt}--\ref{sec:AnSol} computes both cues as differentiable functions of the source direction $(\theta, \phi)$, the frequency $f$, and all geometric and material parameters.

We evaluate the proposed model through a sequence of numerical experiments. We first examine the HRTF, ILD, and ITD patterns produced by the geometry to confirm that they exhibit physically expected directional dependence. We then demonstrate gradient-based source localization from binaural cues, assess robustness to measurement noise across a grid of source directions, evaluate beamforming performance metrics, and demonstrate moving-source tracking with an EKF.
\subsection{Results}
Throughout this section, the exterior medium is water with density $\rho_o = 1000~\mathrm{kg/m^3}$ and sound speed $c_o = 1500~\mathrm{m/s}$, and the interior medium has density $\rho_i = 920~\mathrm{kg/m^3}$ and sound speed $c_i = 1420~\mathrm{m/s}$, representative of biological soft tissue~\cite{Koopman2006}. The semi-transparent outer sphere has radius $a_1 = 0.2~\mathrm{m}$, and the two rigid inner spheres each have radius $a_2 = a_3 = 0.1~\mathrm{m}$. Because the impedance contrast between the interior and exterior media is small ($\rho_i c_i / \rho_o c_o \approx 0.87$), the semi-transparent outer sphere is nearly transparent, and the scattering is dominated by the two rigid inner spheres.

\subsubsection{Two-Rigid-Sphere HRTF}

\begin{figure}[htb]
    \centering
\includegraphics[width=0.75\linewidth]{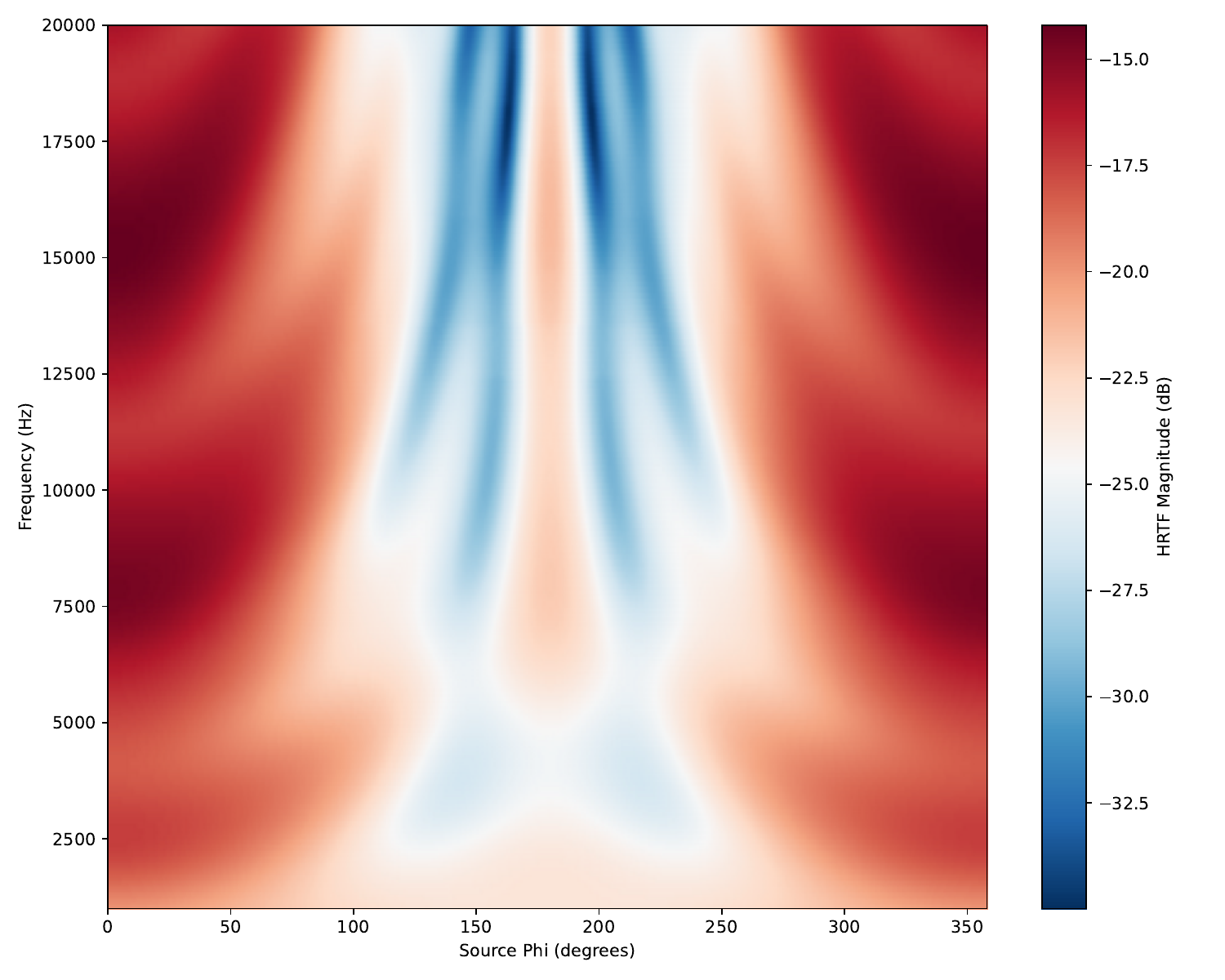}
\caption{Three Spheres HRTF vs Source Phi (degrees) and Frequency (Hz)}
    \label{2-3sph-HRTF}
\end{figure}

Fig.~\ref{2-3sph-HRTF} shows the HRTF of the proposed existing two rigid spheres model in~\cite{Gumerov2002-np}, which here in our study represents the HRTF of two rigid spheres inside a semi-transparent fatty spherical environment when the exterior medium is water. We used this model in our study since it is more accurate, and the HRTFs of the two rigid spheres and the actual model have approximately the same overall structural patterns and magnitude ranges as a function of the frequency and source angle $\phi$. This is because of the fact that in two rigid spheres inside a semi-transparent sphere model, the fat medium has acoustic properties similar to the acoustic properties of the surrounding water. 

\subsubsection{Binaural Cues}

\begin{figure}[!t]
    \centering
    \subfloat[ILD for Symmetrical Microphone sensor positions.]{
\includegraphics[width=8cm, height=6cm]{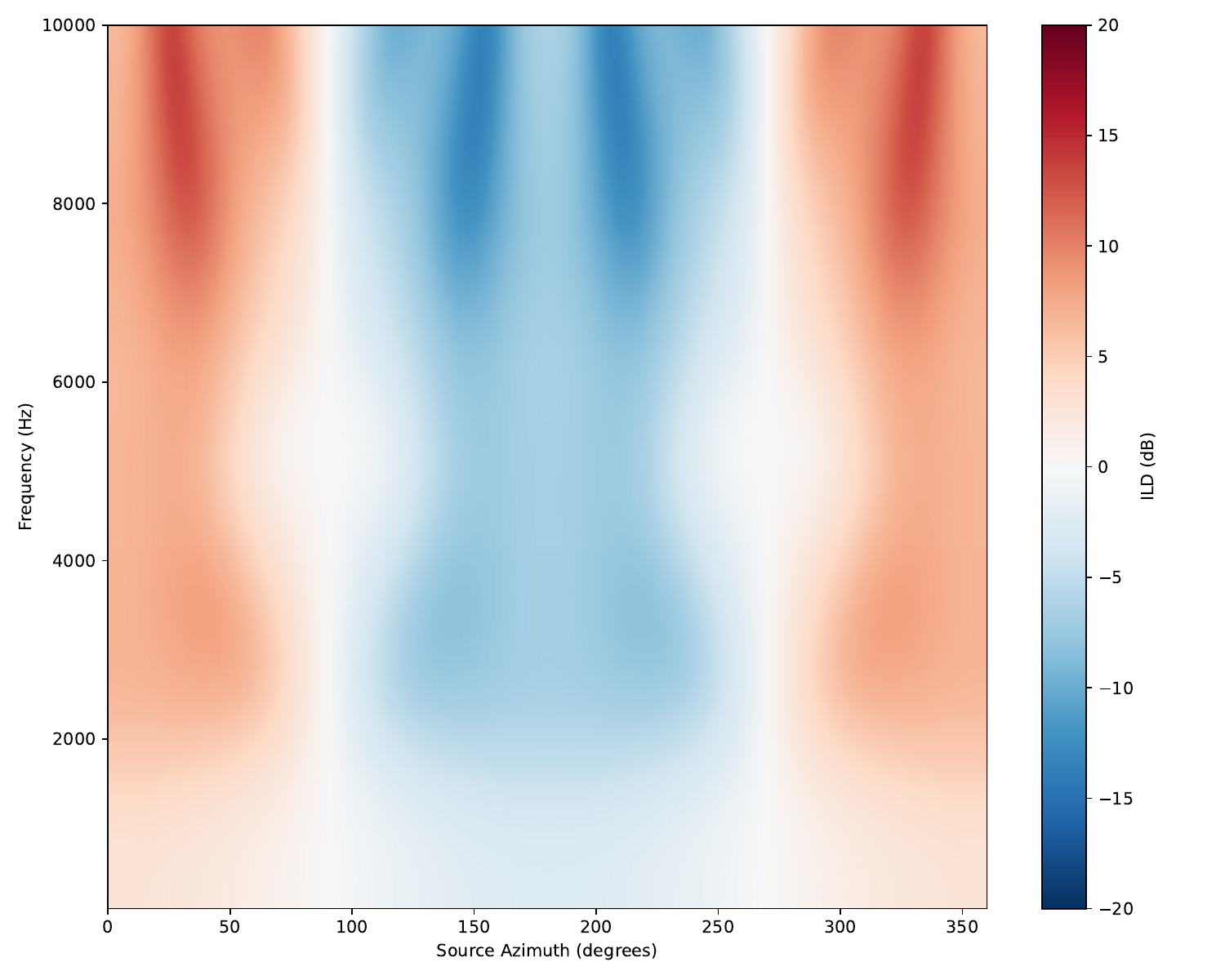}
    \label{fig:ild1}}
    \subfloat[ILD for Asymmetrical Microphone sensor positions.]{
\includegraphics[width=8cm, height=6cm]{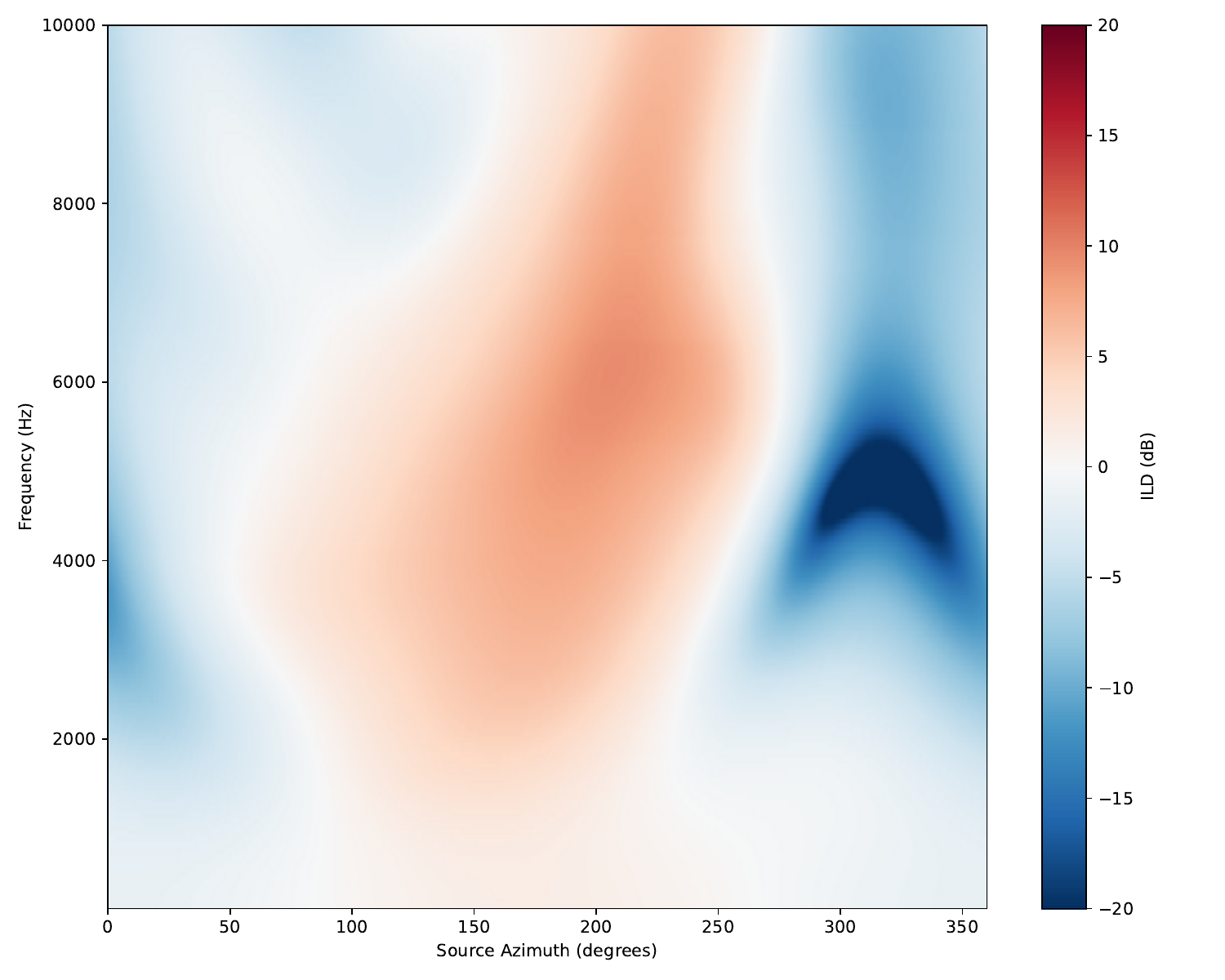}\label{fig:ild2}}
    \caption{ILD vs Source $\phi$ (degrees) and Frequency (Hz).}
    \label{ILD}
\end{figure}

In Fig.~\ref{ILD}, subplot(a) illustrates the variations of ILD with respect to the source azimuth angle and frequency for the symmetrical sensors, while subplot(b) show their corresponding changes for the asymmetrical sensors. From Fig.~\ref{ILD}, several important spatial patterns can be observed. The ILD in subplots (a) and (b) exhibit smooth variations versus azimuth for the lower frequencies ($f \le 1500$). We can observe that when the source is fixed at $\theta = \pi/2$, and the azimuth $\phi$ varies from $0^\circ$ to $360^\circ$, the ILD attains its maximum values at the ends and a minimum around $\phi = 180^\circ$ or the left sensor azimuthal angle. This behavior occurs because the sound source is closer to the right sensor at those azimuths, resulting in receiving a higher pressure and hence a higher ILD. 

\begin{figure}[!t]
    \centering
    \subfloat[ITD for Symmetrical Microphone sensor positions.]{
        \includegraphics[width=8cm, height=6cm]{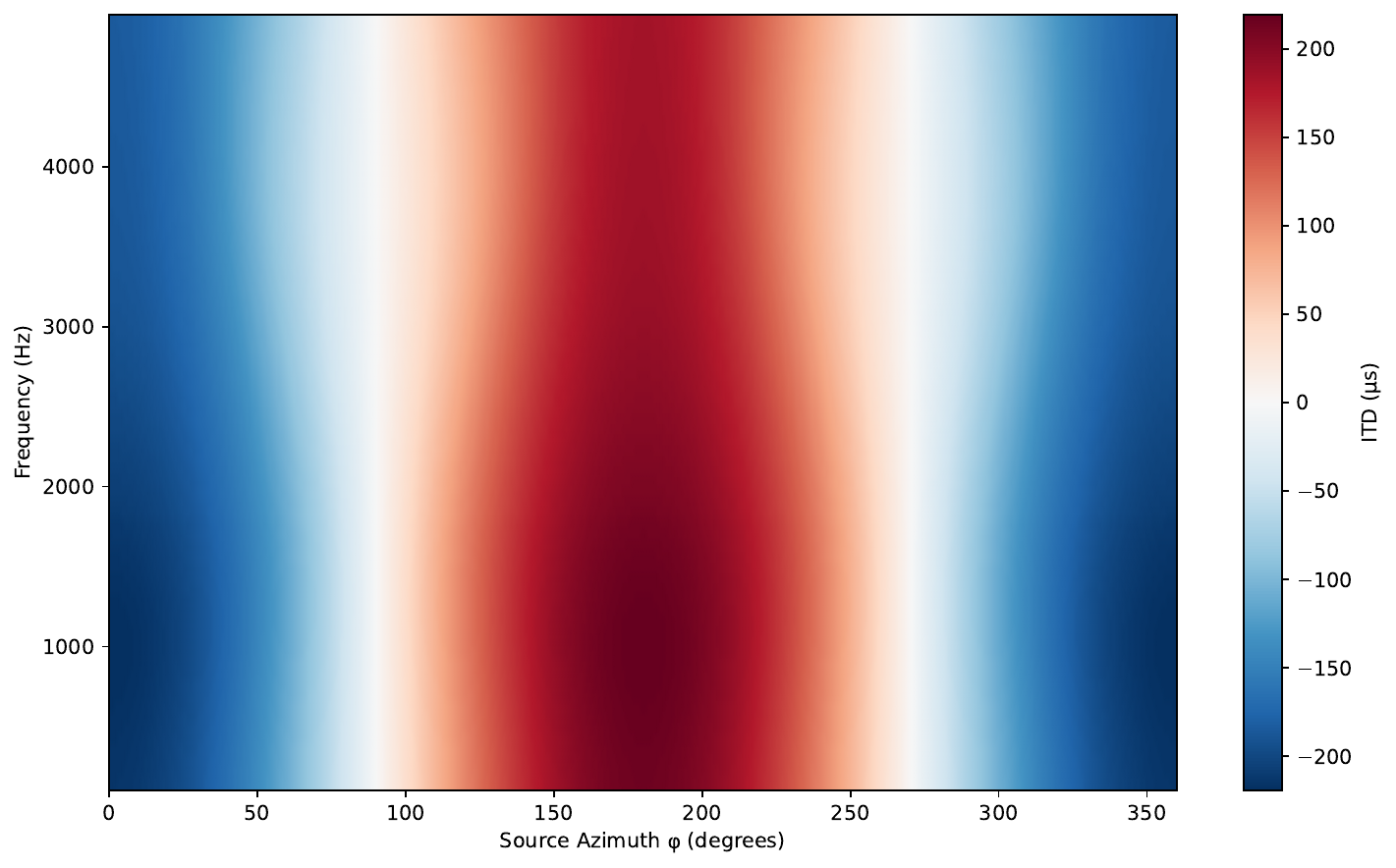}
        \label{fig:itd1}
    }
    \subfloat[ITD for Asymmetrical Microphone sensor positions.]{
        \includegraphics[width=8cm, height=6cm]{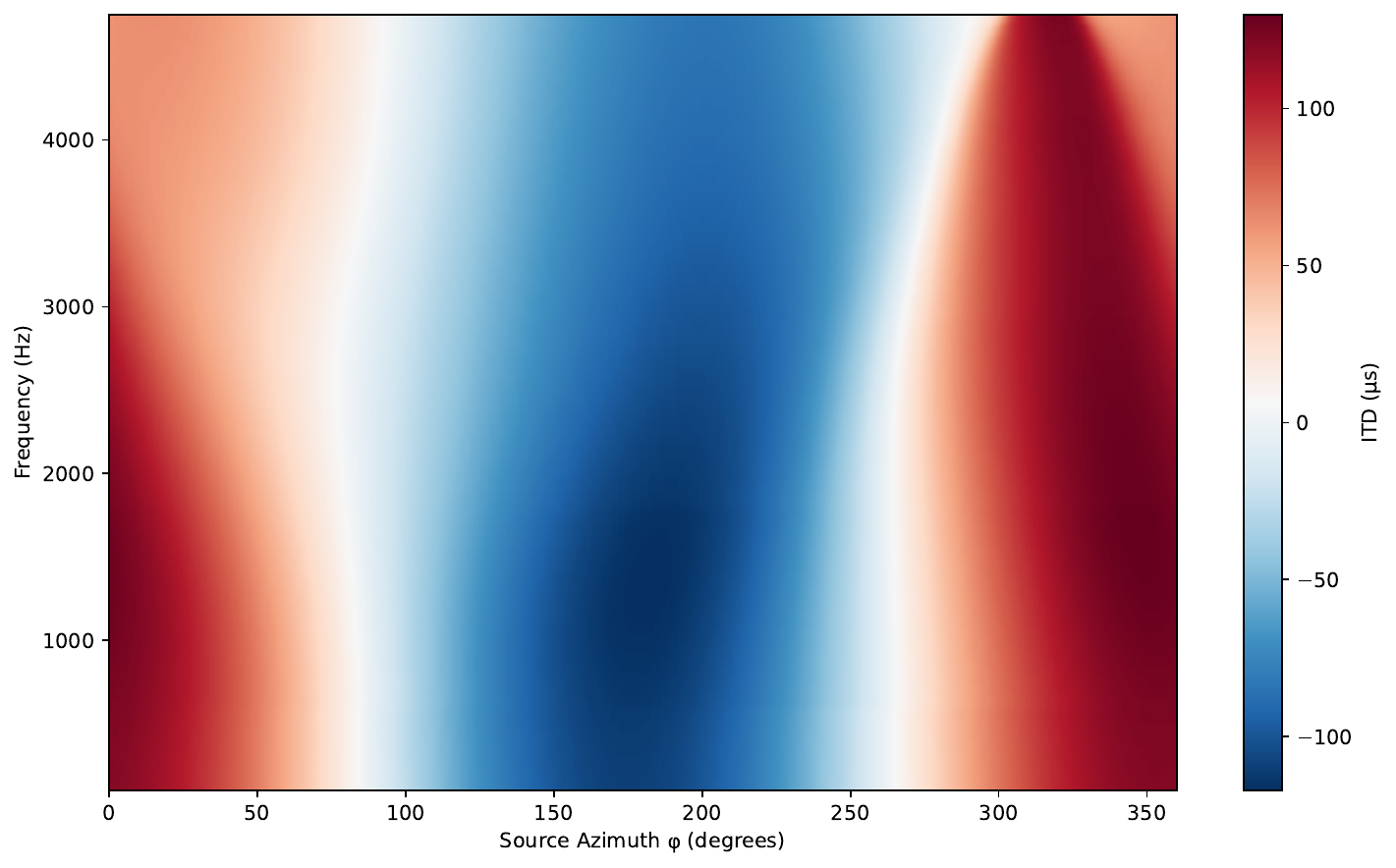}
        \label{fig:itd2}
    }
    \caption{ITD vs source azimuth $\phi$ and frequency (Hz).}
    \label{ITD}
\end{figure}

Fig.~\ref{ITD} shows the ITD variations with respect to the source azimuth angle and frequency. The subplots (a) and (b) demonstrate approximately sinusoidal variations in lower frequencies and changes more smoothly in higher frequencies, with maximum delays occurring when the source is positioned near the sensors and minimal delays at the median plane.  We can observe that when the source is fixed at $\theta = \pi/2$, and the azimuth $\phi$ varies from $0^\circ$ to $360^\circ$, the ITD attains its minimum values at the ends and a maximum around $\phi = 180^\circ$. This behavior occurs because the sound source is closer to the right sensor at smaller azimuths, resulting in a shorter propagation time and hence a smaller ITD. For horizontal sources near the sensors, specifically in the symmetrical case, larger ITD values are consistent with theoretical findings for the spherical scattering model. These results indicate that the model successfully captures the expected directional dependence of binaural cues.

\vspace{1.0em}
We can see that our proposed model generates physically realistic binaural cues that exhibit the correct directional dependence, symmetry properties, and frequency trends expected from underwater scattering. These patterns indicate that the model captures the essential mechanisms by which dolphins extract spatial information. These ILD and ITD characteristics from the HRTF and forward model enable source localization and source tracking algorithms developed in later sections.

\section{Source Localization}
\label{sec:localization}
In this section, we formulate the source localization problem using the proposed scattering model. The goal is to estimate the source direction from received binaural spatial cues. We consider the localization as a continuous optimization problem, where the predicted interaural level and time differences are matched to the measured ones.
\subsection{Definitions and Formulas}
We formulate source localization as a continuous optimization problem. Given a set of observed binaural cues $\mathrm{ILD}^\star(f)$ and $\mathrm{ITD}^\star(f)$ measured at the two sensors, we seek the source direction $(\theta, \phi)$ whose predicted cues best match the observations:
\begin{equation}
\label{eq:optimization_basic}
\min_{\theta,\,\phi}\; \mathcal{L}(\theta, \phi)
\quad \text{s.t.} \quad 
\theta \in [0, \pi], \;
\phi \in [0, 2\pi),
\end{equation}
where the loss function combines the squared ILD and ITD errors across a set of analysis frequencies $\mathcal{F}$:

\begin{equation}
\begin{aligned}
\mathcal{L}(\theta, \phi)
&= \sum_{f \in \mathcal{F}}
\Big[ \big(\overline{\mathrm{ILD}}(f;\theta,\phi) - \overline{\mathrm{ILD}}^\star(f) \big)^2 \\
&\quad + \big(\overline{\mathrm{ITD}}(f;\theta,\phi) - \overline{\mathrm{ITD}}^\star(f)\big)^2 \Big].
\end{aligned}
\label{eq:joint_loss}
\end{equation}
Here, the overline denotes normalized quantities: since the ILD (in dB) and ITD (in seconds) have different magnitudes, both are normalized to comparable scales before summation. Specifically, each cue is divided by its range across the frequency band, so that neither term dominates the loss.

Another metric used to measure the performance of our optimization framework is the angular error between the estimated and true source directions on the unit sphere~\cite{GumerovDuraiswami2004}, and is calculated as following
\begin{equation}
\varepsilon
=
\arccos \!\left(
\frac{4\pi}{3}
\operatorname{Re}
\sum_{s=-1}^{1}
Y_1^{s}(\widehat{\theta},\widehat{\phi})
\,
\overline{Y_1^{s}(\theta^\ast,\phi^\ast)}
\right).
\end{equation}
where $(\widehat{\theta},\widehat{\phi})$ is the estimated direction and 
$(\theta^\ast,\phi^\ast)$ is the true direction and overline denotes complex conjugation.

Because the forward model is implemented in JAX and is fully differentiable, the gradients $\partial \mathcal{L}/\partial \theta$ and $\partial \mathcal{L}/\partial \phi$ are computed exactly via automatic differentiation. We minimize the loss using the Adam optimizer~\cite{adam}, which performs gradient descent with adaptive learning rates. This formulation treats localization as a continuous inverse problem rather than a discrete grid search, with the angular resolution limited only by the optimizer's convergence tolerance rather than by a predefined grid spacing.

To avoid front--back ambiguity in the binaural cues, we place the left and right sensors at slightly asymmetric positions on the sphere surface, breaking the bilateral symmetry that would otherwise produce identical cue patterns for symmetric source pairs.

\subsection{Source Localization: Single Direction}
\label{sec:localization_single}

We first illustrate the localization procedure for a single source direction. The true source is placed at $(\theta^\ast, \phi^\ast) = (2.13,\; 1.10)$~rad. The joint ILD--ITD loss function~\eqref{eq:joint_loss} is minimized using the Adam optimizer with a learning rate of $0.02$, starting from the initial guess $(\theta_0, \phi_0) = (\pi/4,\; \pi/2)$.

\begin{figure}[htb]
    \centering
    \includegraphics[width=0.9\textwidth]{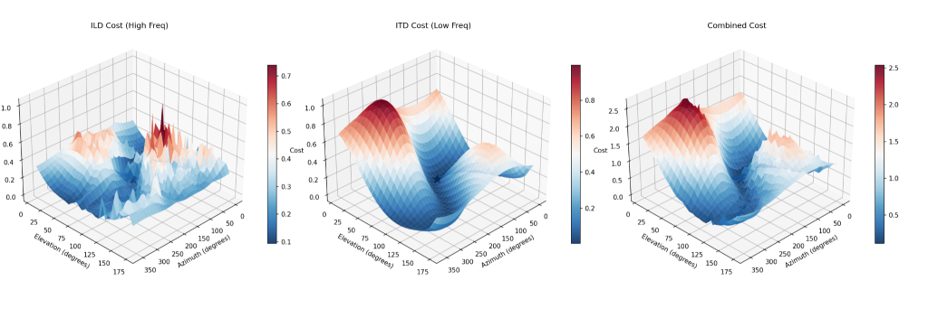}
    \caption{Loss landscape for the joint ILD--ITD objective as a function of source elevation and azimuth. Left: ILD component. Center: ITD component. Right: combined loss showing a single basin of attraction.}
    \label{loss_landscape}
\end{figure}

Fig.~\ref{loss_landscape} shows the loss landscape as a function of elevation and azimuth. The joint loss exhibits a single basin of attraction centered on the true source direction for this configuration. The ILD and ITD components contribute complementary information: the ILD loss has relatively broad contours, while the ITD loss provides sharper localization, and their combination yields a well-defined minimum.

\begin{figure}[htb]
  \centering
  \subfloat[Angular error versus iteration]{%
    \includegraphics[width=0.65\linewidth]{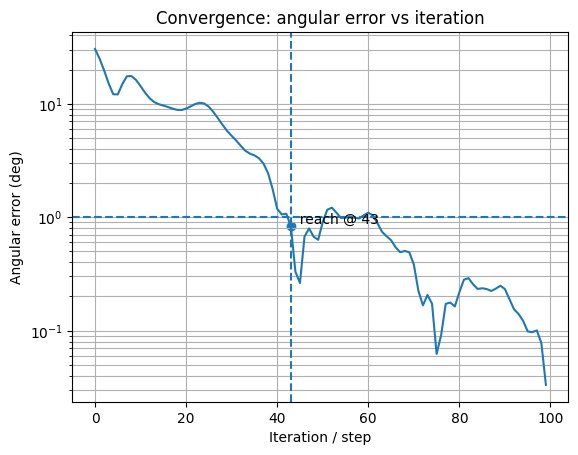}%
    \label{number_steps}%
  }
  \vspace{0.4em}
  \subfloat[Loss and parameter trajectories]{%
    \includegraphics[width=0.85\columnwidth]{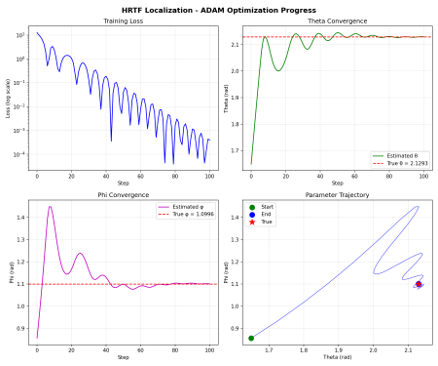}%
    \label{loss_param_traj}%
  }
  \caption{Convergence of gradient-based localization for a single source at $(\theta^\ast, \phi^\ast) = (2.13,\; 1.10)$~rad. (a)~Angular error versus optimizer iteration. (b)~Loss value and estimated $(\theta, \phi)$ versus iteration.}
  \label{fig:source_localization_optimization}
\end{figure}

Fig.~\ref{fig:source_localization_optimization} shows the convergence behavior. The angular error decreases monotonically, falling below $1^\circ$ by iteration~43 and reaching a final value of approximately $0.03^\circ$ (Fig.~\ref{number_steps}). The loss and parameter trajectories (Fig.~\ref{loss_param_traj}) confirm smooth convergence toward the true source direction, with a final loss of $\mathcal{L} = 1.37 \times 10^{-3}$.

To assess sensitivity to initialization, we repeated the optimization from two additional starting points. From $(\theta_0, \phi_0) = (0.57,\; 0.29)$, the optimizer converged to a final angular error of $0.66^\circ$ in 72 iterations. From $(\theta_0, \phi_0) = (2.57,\; 1.29)$, it converged to $0.19^\circ$ in 69 iterations. In all cases, the optimizer reached the vicinity of the true source, though convergence speed and final accuracy depend on the initial distance from the solution.

\subsection{Source Localization: Robustness Across Directions and Noise Levels}
\label{sec:localization_sweep}

To systematically evaluate the localization framework, we tested it across a grid of 32 source directions (4 elevations $\times$ 8 azimuths: $\theta \in \{60^\circ, 90^\circ, 120^\circ, 150^\circ\}$ and $\phi \in \{0^\circ, 45^\circ, \ldots, 315^\circ\}$) at three cue-level signal-to-noise ratios (SNR): 20, 10, and 0~dB. At each direction and SNR, 20 independent noise realizations were tested, for a total of $32 \times 3 \times 20 = 1920$ localization trials.

Noise was applied directly to the ILD and ITD cues. For a given SNR, the noise standard deviation was set proportional to the RMS of the clean cue:
\begin{equation}
\sigma_{\mathrm{cue}} = \frac{\mathrm{RMS}(\mathrm{cue}_{\mathrm{clean}})}{10^{\mathrm{SNR}/20}}.
\label{eq:cue_noise}
\end{equation}
Each trial used a multi-start Adam optimizer with four initializations distributed across the sphere, retaining the result with the lowest final loss. The optimizer ran for a maximum of 100 iterations with early stopping after 30 non-improving steps.

Table~\ref{tab:localization_sweep} summarizes the results. At high SNR (20~dB), the mean angular error is low and the vast majority of trials converge to within $5^\circ$ of the true direction. Performance degrades gracefully as noise increases: at 10~dB, the median error remains moderate, and at 0~dB --- where the noise standard deviation equals the cue RMS --- the optimizer still localizes the source to within $10^\circ$ in most trials, though the fraction of sub-$5^\circ$ results decreases.

\begin{table}[t]
\centering
\caption{Localization performance across 32 directions 
and 20 noise realizations per condition.}
\label{tab:localization_sweep}
\small
\begin{tabular}{c c c c c}
\hline
\textbf{SNR} & \textbf{Mean err.} & \textbf{Median err.} 
  & \textbf{Frac.} & \textbf{Frac.} \\
(dB) & (deg) & (deg) & $<5^\circ$ & $<10^\circ$ \\
\hline
20  & 10.70  & 9.07  & 0.666 & 0.803 \\
10  & 15.25  & 12.66 & 0.391 & 0.631 \\
 0  & 33.85  & 26.57 & 0.097 & 0.267 \\
\hline
\end{tabular}
\end{table}

These results confirm that the differentiable forward model enables reliable gradient-based localization across a range of source directions and noise conditions, with performance scaling predictably with SNR.

\subsection{Robustness to Noise: Detailed Example}
\label{sec:noise_robustness}

\begin{figure}[t]
  \centering
  \subfloat[Angular error versus iteration]{%
    \includegraphics[width=0.65\linewidth]{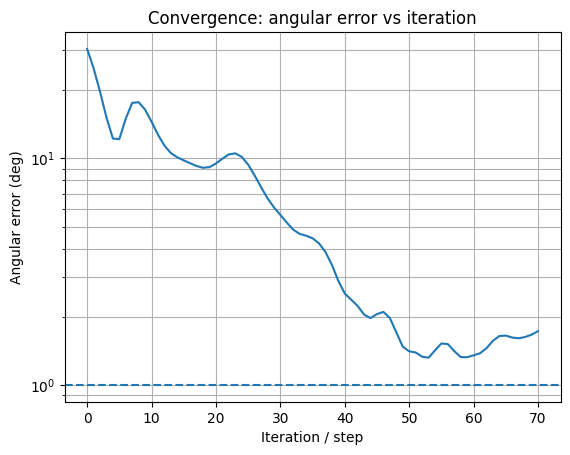}%
    \label{number_step3}%
  }
  \vspace{0.4em}
  \subfloat[Loss and parameter trajectories]{%
    \includegraphics[width=0.85\columnwidth]{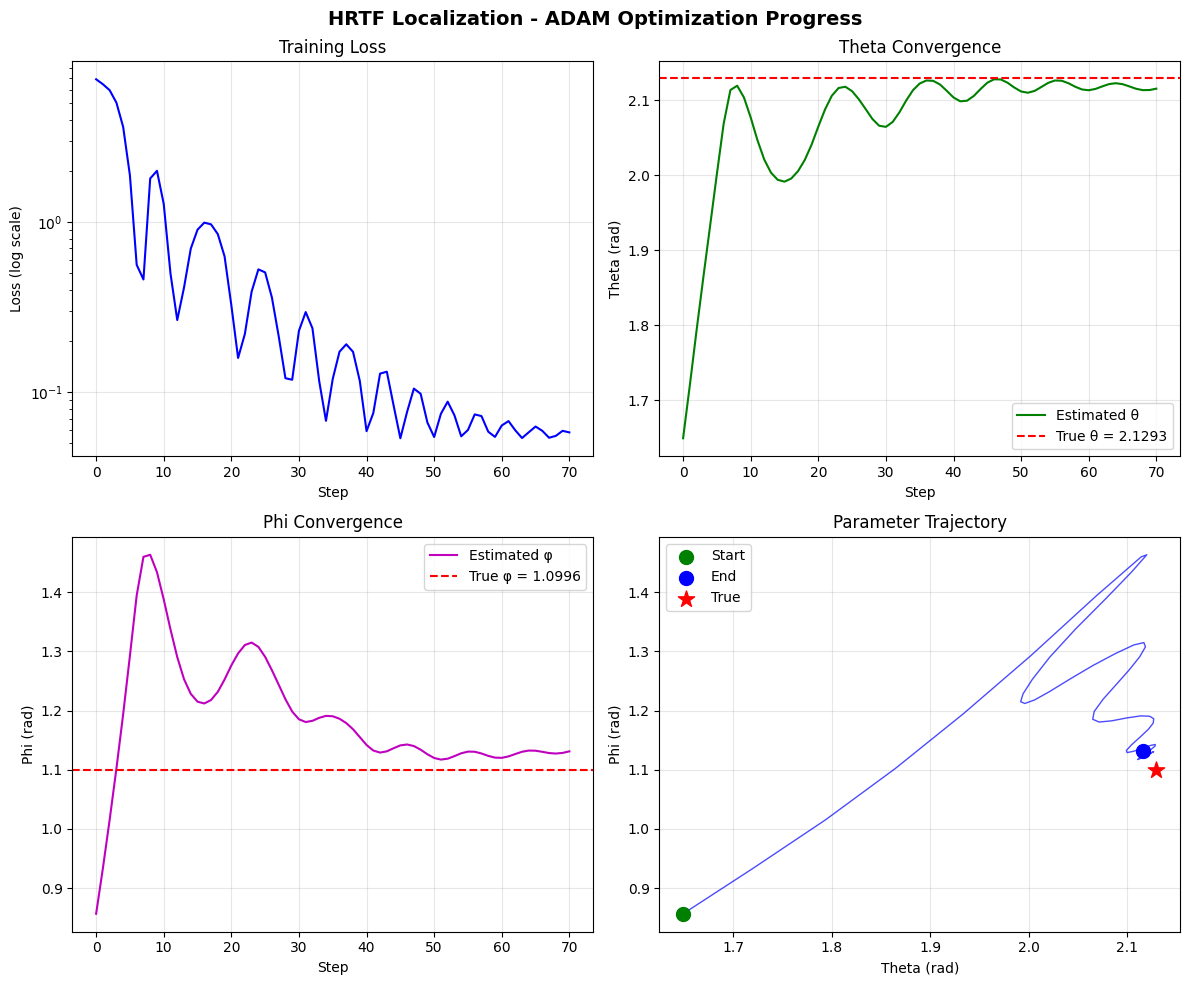}%
    \label{loss_param_traj3}%
  }
  \caption{Localization under measurement noise ($\sigma_{\mathrm{ILD}} = 0.5$~dB, $\sigma_{\mathrm{ITD}} = 10~\mu$s). (a)~Angular error versus iteration. (b)~Loss and parameter trajectories.}
  \label{fig:source_localization_optimization3}
\end{figure}

To illustrate the effect of noise on a single source direction, Fig.~\ref{fig:source_localization_optimization3} shows the convergence behavior under additive Gaussian noise with $\sigma_{\mathrm{ILD}} = 0.5~\mathrm{dB}$ and $\sigma_{\mathrm{ITD}} = 10~\mu\mathrm{s}$. The optimizer converges to a final angular error of $1.73^\circ$, with component-wise errors of $0.78^\circ$ in elevation and $1.81^\circ$ in azimuth. The convergence trajectory remains smooth, though the final loss ($5.80 \times 10^{-2}$) is higher than in the noiseless case, reflecting the irreducible mismatch introduced by the measurement noise.

\section{Spatial Beamforming Using Matched-filter}
Although the primary focus of this work is on spatial cue generation and source localization, microphone sensing systems often employ beamforming to improve signal quality in real-world applications. Therefore, we evaluate the beamforming performance of the proposed microphone array using a matched filter and present its spatial filtering behavior using classical metrics such as White Noise Gain (WNG) and Directivity Index (DI). These metrics measure the robustness of the beamformer to sensor noise and its spatial selectivity, respectively~\cite{vantrees2002,urick1983}.
\subsection{Definition and Formulas}
Assume the complex frequency-domain steering vector corresponding to a source at direction $\Omega = (\theta,\phi)$ be defined as
\begin{equation}
\mathbf{h}(f,\Omega) =
\begin{bmatrix}
H_L(f,\Omega) \\
H_R(f,\Omega)
\end{bmatrix},
\end{equation}
where $H_L$ and $H_R$ denote the HRTFs at the left and right microphone sensors computed using the proposed scattering model, respectively.

For a look direction $\Omega_0$, the matched filter combines the signal received from the two channels with the following weights,
\begin{equation}
\mathbf{w}(f)
=
\frac{\mathbf{h}_0(f)}
{\mathbf{h}_0^{H}(f)\,\mathbf{h}_0(f)}
\end{equation}
This normalization ensures a distortionless response in the look direction, i.e., $\mathbf{w}^H(f)\mathbf{h}_0(f)=1$, where the steering vector at the look direction is
\begin{equation}
\mathbf{h}_0(f)
=
\begin{bmatrix}
H_L(f,\Omega_0) \\
H_R(f,\Omega_0)
\end{bmatrix}.
\end{equation}
The beamformer output is then
\begin{equation}
Y(f) = \mathbf{w}^H(f)\mathbf{X}(f),
\end{equation}
where $\mathbf{X}(f)$ is the vector of the signals received by two microphones.

\paragraph{WNG} measures the sensitivity of the beamformer to spatially uncorrelated white sensor noise. The normalized WNG is defined as
\begin{equation}
\mathrm{WNG}(f) =
\frac{1}{\mathbf{w}^H(f)\mathbf{w}(f)}.
\end{equation}

\paragraph{DF} evaluates the spatial selectivity of the beamformer, defined as the ratio between the received beam power in the look direction and the average beam power over all directions:
\begin{equation}
\mathrm{DF}(f) =
\frac{\left| \mathbf{w}^H(f)\mathbf{h}(f,\Omega_0) \right|^2}
{\frac{1}{4\pi}\int_{4\pi}
\left| \mathbf{w}^H(f)\mathbf{h}(f,\Omega) \right|^2
\, d\Omega }.
\end{equation}
The spherical integral over all directions in the above equation is approximated by using a discretized direction grid with angle weighting. The Directivity Index (DI) is the expression of DF in decibels as
\begin{equation}
\mathrm{DI}(f) = 10 \log_{10} \left( \mathrm{DF}(f) \right).
\end{equation}
\subsection{Results}

Using the matched-filter beamformer steered toward the look direction 
$(\theta, \phi) = (2.1293, 1.0996)$ rad, we first verified distortionless response across all 
evaluated frequencies, with $|\mathbf{w}^H \mathbf{h}_0| = 1.0$, confirming proper normalization. The WNG remained positive over the full band, ranging from 
$3.35\,\mathrm{dB}$ to $5.96\,\mathrm{dB}$ (mean $\approx 5.21\,\mathrm{dB}$), 
indicating stable noise amplification behavior without excessive sensitivity to sensor noise.

\begin{figure}[t]
    \centering 
    \vspace{0.4em}
    \includegraphics[width=0.69\linewidth]{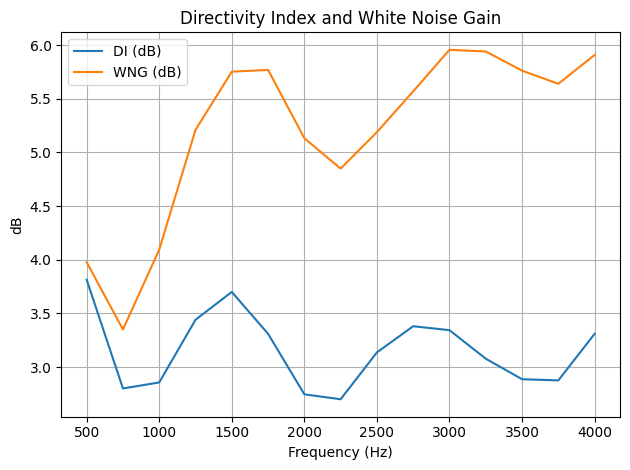}
    \caption{WNG and DI vs. Frequency}
    \label{wng_di_frq}
\end{figure}

The DF and DI were computed over a $162$-direction spherical grid. The DF varied between $1.86$ and $2.41$, 
corresponding to a DI range of $2.70\,\mathrm{dB}$ to $3.81\,\mathrm{dB}$ 
(with mean DI $\approx 3.16\,\mathrm{dB}$). 
Fig.~\ref{wng_di_frq} demonstrates the range of the WNG and DI versus frequency in dB.
\section{Tracking a Moving Source}
In this section, we evaluate the temporal beamforming performance of this proposed two-microphone system. We perform temporal filtering by applying a Kalman filter to understand and measure the performance of the proposed system.
\subsection{Definition and Formulas}
To enable continuous tracking of a moving acoustic source, we employ an EKF with a constant-velocity state model~\cite{kalman1960,simon2006,barshalom2001}. The state vector is defined as
\begin{equation}
\mathbf{x}_t =
\begin{bmatrix}
\theta_t & \phi_t & \dot{\theta}_t & \dot{\phi}_t
\end{bmatrix}^T ,
\end{equation}
where $\theta_t$ and $\phi_t$ denote the elevation and azimuth of the source, respectively.

Considering constant angular velocity with sampling interval $\Delta t$, the motion model is
\begin{equation}
\mathbf{x}_t = \mathbf{F}\mathbf{x}_{t-1} + \mathbf{w}_t ,
\end{equation}
with
\begin{equation}
\mathbf{F}
=
\begin{bmatrix}
1 & 0 & \Delta t & 0 \\
0 & 1 & 0 & \Delta t \\
0 & 0 & 1 & 0 \\
0 & 0 & 0 & 1
\end{bmatrix},
\end{equation}
and process noise $\mathbf{w}_t \sim \mathcal{N}(\mathbf{0},\mathbf{Q})$. Further details on the process noise model are provided in Appendix~\ref{app:Ang_Acc}.

The nonlinear measurement model is
\begin{equation}
\mathbf{z}_t = h(\mathbf{x}_t) + \mathbf{v}_t ,
\end{equation}
where $\mathbf{z}_t$ stacks the measured ILD and ITD cues across frequency bins, forming a high-dimensional observation vector. $h(\cdot)$ denotes the physics-based forward acoustic model, and $\mathbf{v}_t \sim \mathcal{N}(\mathbf{0},\mathbf{R})$ is measurement noise.

At each time step, the EKF performs prediction
\begin{align}
\hat{\mathbf{x}}_{t|t-1} &= \mathbf{F}\hat{\mathbf{x}}_{t-1|t-1}, \\
\mathbf{P}_{t|t-1} &= \mathbf{F}\mathbf{P}_{t-1|t-1}\mathbf{F}^T + \mathbf{Q},
\end{align}
Here, $\mathbf{P}_{t|t}$ denotes the state estimation error covariance matrix, which quantifies the uncertainty in the angular position and velocity estimates.

Since the physics-based acoustic measurement function $h(\mathbf{x})$ is nonlinear in the source angles, the EKF linearizes it via a first-order Taylor expansion around the predicted state~\cite{simon2006}. The Jacobian $\mathbf{H}_t$ captures the local sensitivity of the ILD and ITD cues to small angular perturbations, thereby enabling recursive uncertainty propagation.

\begin{equation}
\mathbf{H}_t =
\left.
\frac{\partial h(\mathbf{x})}{\partial \mathbf{x}}
\right|_{\hat{\mathbf{x}}_{t|t-1}},
\end{equation}
The Jacobian is obtained from the differentiable acoustic model, enabling consistent linearization of the measurement function, which yields
\begin{align}
\mathbf{K}_t &= \mathbf{P}_{t|t-1}\mathbf{H}_t^T
\left(\mathbf{H}_t \mathbf{P}_{t|t-1} \mathbf{H}_t^T + \mathbf{R}\right)^{-1}, \\
\hat{\mathbf{x}}_{t|t} &= \hat{\mathbf{x}}_{t|t-1}
+ \mathbf{K}_t \big( \mathbf{z}_t - h(\hat{\mathbf{x}}_{t|t-1}) \big), \\
\mathbf{P}_{t|t} &= (\mathbf{I} - \mathbf{K}_t \mathbf{H}_t)\mathbf{P}_{t|t-1}.
\end{align}
The detailed derivation of the EKF equations is provided in Appendix~\ref{app:ekf_derivation}.

\begin{figure}[t]
\centering
 \vspace{0.2em}
{\includegraphics[width=0.65\linewidth]{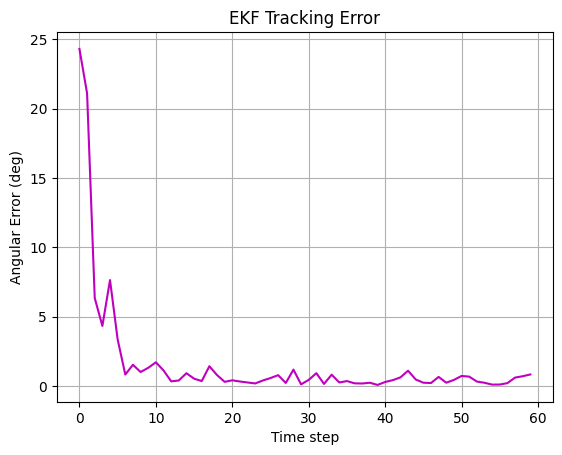}}
\caption{EKF angular tracking error over time.}
\label{fig:ekf_ang_err}
\end{figure}

\begin{figure}[htb]
\centering

\includegraphics[width=0.65\linewidth]{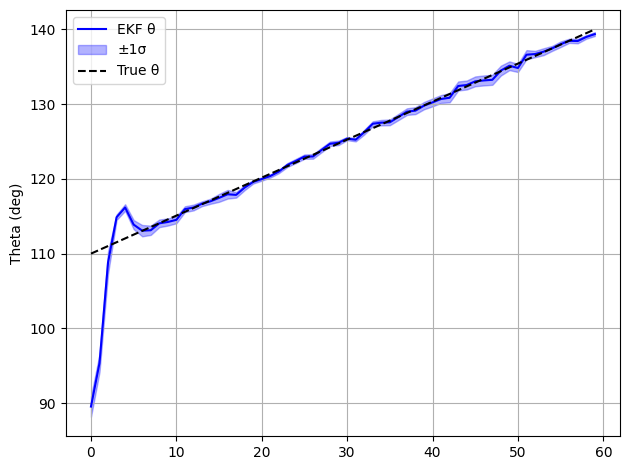}

\vspace{0.5em}

\includegraphics[width=0.65\linewidth]{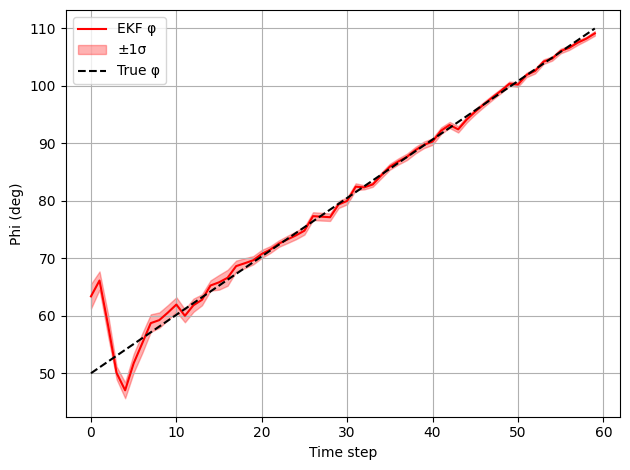}

\caption{EKF tracking results for elevation (top) and azimuth (bottom).}
\label{fig:ekf_tracking}
\end{figure}
\subsection{Results}

We assume a moving sound source over $T = 60$ discrete time steps. 
The source trajectory was defined as a smooth linear motion in spherical 
coordinates, with elevation $\theta$ varying from $110^\circ$ to $140^\circ$ 
and azimuth $\phi$ varying from $50^\circ$ to $110^\circ$. 
At each time step, frequency-dependent ILD and ITD cues were generated by our proposed forward model.

Measurement noise was modeled as zero-mean Gaussian noise added 
to ILD and ITD observations. The assumed standard deviations were 
$\sigma_{\mathrm{ILD}} = 0.5\,\mathrm{dB}$ and 
$\sigma_{\mathrm{ITD}} = 10\,\mu\mathrm{s}$, consistent cue uncertainty in real audio sensing systems.

Source tracking was performed using an EKF with a time step $\Delta t = 1.0$. 
Process noise was introduced through angular acceleration terms with 
$\sigma_{\theta,\mathrm{acc}} = 0.03$ and 
$\sigma_{\phi,\mathrm{acc}} = 0.03$, 
allowing smooth but non-deterministic trajectory evolution (look at Appendix~\ref{app:Ang_Acc}). 
The filter was initialized at $(\theta_0, \phi_0)=(2.2, 2.6)$ and recursively estimated 
the source angles from the noisy multi-frequency ILD and ITD measurements.

Fig.~\ref{fig:ekf_ang_err} demonstrates the tracking performance of the proposed EKF. After a few steps from the imperfect initial point, the angular error rapidly decreases as the filter's output converges to the true point. After convergence, it has been shown in Fig.~\ref{fig:ekf_tracking} that the tracking error remains below $1^\circ$ for most of the trajectory, as the $\pm 1\sigma$ confidence bands shrink after the transient phase, indicating increasing estimation certainty. These results demonstrate having a differentiable forward model enables the EKF to linearize the nonlinear forward model and achieve accurate and stable source tracking.

\section{Discussion}
Unlike previous works, our work  employs  modeling of scattering to create a device that allows spatial cues to be generated that enable accurate localization and stable tracking with a minimal two-sensor array. By modeling the acoustic scatter field, the framework captures binaural spatial cues that would be difficult to produce under free-field assumptions. 
Future work may explore more sophisticated motion models, integration with real-world measurements, and evaluation in complex acoustic environments with reverberation and multiple sources. While the proposed framework considers a minimal number of sensors, leveraging additional sensors could increase robustness. Additionally, investigating adaptive or learning-based cue fusion strategies may improve performance under challenging conditions. 

\section{Conclusion}
\noindent
We study the performance of the two-sensor system with a multi-spherical scattering model, inspired by the internal anatomical features of the underwater mammals. Using the Multipole Expansion method, we investigate how dolphins generate and utilize spatial cues for underwater navigation and localization. Our study introduces an analytical and numerical framework for determining the HRTF at the left and right sensors of this multi-spherical model. By solving the acoustic scattering problem, we obtain consistent HRTF, ITD, and ILD patterns that characterize the spatial cues used in localization, beamforming, and tracking. The agreement between the analytical and numerical results validates the accuracy of the proposed model and its physical realism. This work provides a mathematically and physically interpretable holistic foundation for understanding the spatial mechanisms of the scattering systems including in an underwater setting. 
\clearpage
\appendix
\section*{Appendix}
\section{Coupled linear system}
\label{linear_system}
Equations~\eqref{eq:bc1_vel}, \eqref{eq:bc1_pres}, \eqref{eq:bc2}, and~\eqref{eq:bc3} form a coupled infinite-dimensional linear system in the unknown coefficients $A^{(1)}$, $B^{(1)}$, $C^{(1)}$, and $D^{(3)}$. For numerical solution, we truncate the expansions at degree $p-1$, retaining modes $l = 0, 1, \ldots, p-1$. Following~\cite{Gumerov2002-np}, convergence to within $10^{-3}$ is achieved for $p \ge 3 k_i a_1$.

To express the truncated system in matrix-vector form, we define the following quantities. The translation matrices for each order $s$ are
\begin{equation}
\resizebox{0.9\columnwidth}{!}{$
(X|R)^s(\mathbf{t})
= \begin{pmatrix}
\scriptstyle (X|R)_{|s|,|s|}^s   & \scriptstyle (X|R)_{|s|,|s|+1}^s   & \cdots & \scriptstyle (X|R)_{|s|,p-1}^s   \\
\scriptstyle (X|R)_{|s|+1,|s|}^s & \scriptstyle (X|R)_{|s|+1,|s|+1}^s & \cdots & \scriptstyle (X|R)_{|s|+1,p-1}^s \\
\vdots               & \vdots                  & \ddots & \vdots               \\
\scriptstyle (X|R)_{p-1,|s|}^s    & \scriptstyle (X|R)_{p-1,|s|+1}^s    & \cdots & \scriptstyle (X|R)_{p-1,p-1}^s
\end{pmatrix}
$}
\label{eq:translation_matrix}
\end{equation}
where $X \in \{S, R\}$, $s = 0, \pm 1, \ldots, \pm(p-1)$, and $\mathbf{t}$ denotes the translation vector between sphere centers. The diagonal matrices encoding the local boundary relations on each sphere are
\begin{align}
\boldsymbol{\Lambda}^{(q)s}_f
  &= \operatorname{diag}\!\Big(
       \Lambda_{|s|}^{(q)f},\;
       \Lambda_{|s|+1}^{(q)f},\;
       \ldots,\;
       \Lambda_{p-1}^{(q)f}
     \Big),
\label{eq:Lambda}\\[4pt]
\boldsymbol{\Gamma}^{(q)s}_f
  &= \operatorname{diag}\!\Big(
       \Gamma_{|s|}^{(q)f},\;
       \Gamma_{|s|+1}^{(q)f},\;
       \ldots,\;
       \Gamma_{p-1}^{(q)f}
     \Big),
\label{eq:Gamma}
\end{align}
with entries
\begin{equation}
\Lambda_l^{(q)f}
= \frac{j_l'(k_f a_q)}{h_l'(k_f a_q)},
\qquad
\Gamma_l^{(q)f}
= \frac{j_l(k_f a_q)}{h_l(k_f a_q)},
\label{eq:LambdaGamma_def}
\end{equation}
for $l = |s|, \ldots, p-1$, $q \in \{1, 2, 3\}$, and $f \in \{i, o\}$. These ratios arise from expressing the boundary conditions in terms of the ratio of regular to singular basis functions evaluated on each sphere surface.

The unknown coefficient vectors are
\begin{equation}
\mathbf{X}^{(q)s}
= \begin{pmatrix}
    X_{|s|}^{(q)s} \\[2pt]
    \vdots \\[2pt]
    X_{p-1}^{(q)s}
  \end{pmatrix},
\qquad X \in \{A,B,C,D\},
\label{eq:coeff_vectors}
\end{equation}
and the source terms, which arise from the known incident field coefficients and the transmission conditions on $S_1$, are
\begin{equation}
\resizebox{0.9\columnwidth}{!}{$
\mathbf{E}^{(1)s}
= \begin{pmatrix}
    E_{|s|}^{s}\;\dfrac{j_{|s|}(k_o a_1)}{j_{|s|}(k_i a_1)}
    \\[6pt] \vdots \\[6pt]
    E_{p-1}^{s}\;\dfrac{j_{p-1}(k_o a_1)}{j_{p-1}(k_i a_1)}
  \end{pmatrix},
\qquad
\mathbf{E}^{(2)s}
= \begin{pmatrix}
    E_{|s|}^{s}\;\dfrac{j_{|s|}'(k_o a_1)}{j_{|s|}'(k_i a_1)}
    \\[6pt] \vdots \\[6pt]
    E_{p-1}^{s}\;\dfrac{j_{p-1}'(k_o a_1)}{j_{p-1}'(k_i a_1)}
  \end{pmatrix}
$}
\label{eq:source_vectors}
\end{equation}
where $s = 0, \pm 1, \ldots, \pm(p-1)$.

Assembling these definitions, the coupled system for each order $s$ takes the block-matrix form:
\begin{equation}
\resizebox{0.9\columnwidth}{!}{$
\begin{pmatrix}
-\boldsymbol{\Gamma}^{(1)s}_o & \mathbf{I}
  & \boldsymbol{\Gamma}^{(1)s}_i & (S|R)^s(\mathbf{r}'_{31}) \\[4pt]
-\boldsymbol{\Lambda}^{(1)s}_o & \mathbf{I}
  & \boldsymbol{\Lambda}^{(1)s}_i & (S|R)^s(\mathbf{r}'_{31}) \\[4pt]
\mathbf{0} & \mathbf{I}
  & \boldsymbol{\Gamma}^{(2)s}_i & (S|R)^s(\mathbf{r}'_{31}) \\[4pt]
\mathbf{0} & -(R|R)^s(\mathbf{r}'_{13})
  & -(S|R)^s(\mathbf{r}'_{13})
  & \boldsymbol{\Gamma}^{(3)s}_i
\end{pmatrix}
\begin{pmatrix}
\mathbf{A}^{(1)s} \\[4pt]
\mathbf{B}^{(1)s} \\[4pt]
\mathbf{C}^{(1)s} \\[4pt]
\mathbf{D}^{(3)s}
\end{pmatrix}
=
\begin{pmatrix}
\mathbf{E}^{(1)s} \\[4pt]
\mathbf{E}^{(2)s} \\[4pt]
\mathbf{0} \\[4pt]
\mathbf{0}
\end{pmatrix}
$}
\label{eq:coupled_system}
\end{equation}
Each row block corresponds to a boundary condition: the first two rows enforce the transmission conditions on $S_1$ (pressure and velocity continuity), the third row enforces the Neumann condition on $S_2$, and the fourth enforces it on $S_3$. For each order $s$, the system has $4(p - |s|)$ equations and $4(p - |s|)$ unknowns.

The system~\eqref{eq:coupled_system} is solved independently for each order $s = 0, \pm 1, \ldots, \pm(p-1)$. We use LU decomposition: the system matrix is factorized once per order and the factors are reused across multiple right-hand sides corresponding to different frequencies and incident field configurations. The computational cost per order is $\mathcal{O}\!\big((p - |s|)^3\big)$, dominated by the matrix factorization~\cite{Gumerov2002-np}. 

\section{Noise Modeling via Angular Acceleration}
\label{app:Ang_Acc}
The state-transition model assumes a constant angular velocity between time steps. 
However, smooth variations exist in velocity due to the angular acceleration of the real moving sources. This uncertainty represents process noise and is modeled as angular acceleration~\cite{simon2006}.

Consider the continuous angular motion in time be described by

\begin{equation}
\ddot{\theta}(t) = a_\theta(t), 
\qquad
\ddot{\phi}(t) = a_\phi(t),
\end{equation}
where $a_\theta(t)$ and $a_\phi(t)$ are zero-mean white Gaussian acceleration processes with variances 
$\sigma_{\theta,\mathrm{acc}}^2$ and $\sigma_{\phi,\mathrm{acc}}^2$, respectively. Accelerations are assumed independent across angular dimensions.

Discretizing over sampling interval $\Delta t$ yields the stochastic state update:

\begin{equation}
\mathbf{x}_t = \mathbf{F}\mathbf{x}_{t-1} + \mathbf{w}_t,
\end{equation}
where the process noise covariance matrix $\mathbf{Q}$ is derived from the present acceleration noise and calculated as follows:

\begin{equation}
\mathbf{Q} =
\begin{bmatrix}
\frac{\Delta t^4}{4}\sigma_{\theta,\mathrm{acc}}^2 & 0 & \frac{\Delta t^3}{2}\sigma_{\theta,\mathrm{acc}}^2 & 0 \\
0 & \frac{\Delta t^4}{4}\sigma_{\phi,\mathrm{acc}}^2 & 0 & \frac{\Delta t^3}{2}\sigma_{\phi,\mathrm{acc}}^2 \\
\frac{\Delta t^3}{2}\sigma_{\theta,\mathrm{acc}}^2 & 0 & \Delta t^2\sigma_{\theta,\mathrm{acc}}^2 & 0 \\
0 & \frac{\Delta t^3}{2}\sigma_{\phi,\mathrm{acc}}^2 & 0 & \Delta t^2\sigma_{\phi,\mathrm{acc}}^2
\end{bmatrix}.
\end{equation}
The acceleration variances $\sigma_{\theta,\mathrm{acc}}$ and $\sigma_{\phi,\mathrm{acc}}$ control the flexibility of motion model. 
Larger values permit faster trajectory changes, whereas smaller values enforce smoother motion.

\section{Derivation of the EKF Update Equations}
\label{app:ekf_derivation}
The EKF update equations for tracking the moving sound source are derived as follows~\cite{simon2006}. 

The measurement model is given by
\begin{equation}
\mathbf{z}_t = h(\mathbf{x}_t) + \mathbf{v}_t,
\end{equation}
where $h(\cdot)$ is the nonlinear forward HRTF model that maps source angles to the stacked ILD and ITD vector, and
\[
\mathbf{v}_t \sim \mathcal{N}(\mathbf{0}, \mathbf{R})
\]
is zero-mean Gaussian measurement noise.

At time step $t$, the predicted state and covariance are denoted by
\[
\hat{\mathbf{x}}_{t|t-1}, \qquad \mathbf{P}_{t|t-1}.
\]
Since $h(\mathbf{x})$ is nonlinear in $(\theta,\phi)$, we approximate it using a first-order Taylor expansion around the predicted state:
\begin{equation}
h(\mathbf{x}_t)
\approx
h(\hat{\mathbf{x}}_{t|t-1})
+
\mathbf{H}_t
(\mathbf{x}_t - \hat{\mathbf{x}}_{t|t-1}),
\end{equation}
where the Jacobian matrix is
\begin{equation}
\mathbf{H}_t
=
\left.
\frac{\partial h(\mathbf{x})}{\partial \mathbf{x}}
\right|_{\hat{\mathbf{x}}_{t|t-1}}.
\end{equation}

Define the innovation (measurement residual)
\begin{equation}
\mathbf{y}_t
=
\mathbf{z}_t
-
h(\hat{\mathbf{x}}_{t|t-1}).
\end{equation}
Substituting the linearized model yields
\begin{equation}
\mathbf{y}_t
\approx
\mathbf{H}_t
(\mathbf{x}_t - \hat{\mathbf{x}}_{t|t-1})
+
\mathbf{v}_t.
\end{equation}
This corresponds to a locally linear measurement equation. The innovation covariance is
\begin{equation}
\mathbf{S}_t
=
\mathbf{H}_t \mathbf{P}_{t|t-1} \mathbf{H}_t^T
+
\mathbf{R}.
\end{equation}
The optimal Kalman gain that minimizes the posterior mean-square estimation error is
\begin{equation}
\mathbf{K}_t
=
\mathbf{P}_{t|t-1}\mathbf{H}_t^T
\mathbf{S}_t^{-1}
=
\mathbf{P}_{t|t-1}\mathbf{H}_t^T
\left(
\mathbf{H}_t \mathbf{P}_{t|t-1} \mathbf{H}_t^T
+
\mathbf{R}
\right)^{-1}.
\end{equation}
The posterior state estimate is updated as
\begin{equation}
\hat{\mathbf{x}}_{t|t}
=
\hat{\mathbf{x}}_{t|t-1}
+
\mathbf{K}_t
\left(
\mathbf{z}_t
-
h(\hat{\mathbf{x}}_{t|t-1})
\right).
\end{equation}

The posterior covariance can be written in the simplified form
\begin{equation}
\mathbf{P}_{t|t}
=
(\mathbf{I} - \mathbf{K}_t \mathbf{H}_t)
\mathbf{P}_{t|t-1}.
\end{equation}
For improved numerical stability, the Joseph form may also be used:
\begin{equation}
\mathbf{P}_{t|t}
=
(\mathbf{I} - \mathbf{K}_t \mathbf{H}_t)
\mathbf{P}_{t|t-1}
(\mathbf{I} - \mathbf{K}_t \mathbf{H}_t)^T
+
\mathbf{K}_t \mathbf{R} \mathbf{K}_t^T.
\end{equation}

Note: Because the acoustic measurement function depends only on the angular components $(\theta,\phi)$ of the state vector
\[
\mathbf{x}_t =
\begin{bmatrix}
\theta_t & \phi_t & \dot{\theta}_t & \dot{\phi}_t
\end{bmatrix}^T,
\]
the Jacobian has the block structure
\begin{equation}
\mathbf{H}_t =
\left[
\frac{\partial h}{\partial \theta}
\quad
\frac{\partial h}{\partial \phi}
\quad
\mathbf{0}
\quad
\mathbf{0}
\right]_{\hat{\mathbf{x}}_{t|t-1}}.
\end{equation}

\noindent 

\section*{Author Contributions}

\noindent \textbf{Siminfar Samakoush Galougah}  developed the analytical framework, implemented the primary simulations, conducted the experiments, and led the writing of the manuscript.

\noindent \textbf{Pranav Pulijala} contributed to the development and implementation of the simulation code, debugged the software, conducted some of the experiments, wrote portions of the manuscript, and participated in technical discussions and manuscript revisions.

\noindent \textbf{Ramani Duraiswami} conceived the study, provided conceptual guidance, obtained the funding, and reviewed and edited the manuscript.

\section*{Ethics}

\noindent There weres no real world experiments on human or animals in this study.
\section*{Glossary}
\noindent
\textbf{DF}: Directivity Factor.\\
\textbf{DI}: Directivity Index.\\
\textbf{EKF}: Extended Kalman Filter.\\
\textbf{HRTF}:Head Related Transfer Function.\\
\textbf{ILD}: Interaural Level Difference.\\
\textbf{ITD}: Interaural Time Difference.\\
\textbf{WNG}: White Noise Gain.


\bibliographystyle{unsrt}
\bibliography{sampbib}



%
%
%
%
%
%


\end{document}